\documentclass[]{elsarticle} 
\usepackage[hyphens]{url}
\usepackage{lineno} 
\providecommand{\tightlist}{%
\setlength{\itemsep}{0pt}\setlength{\parskip}{0pt}}

\biboptions{sort&compress} 
\usepackage{graphicx}
\usepackage{booktabs} 

\usepackage[T1]{fontenc}
\usepackage{lmodern}
\usepackage{amssymb,amsmath}
\usepackage{ifxetex,ifluatex}
\usepackage{fixltx2e} 
\IfFileExists{upquote.sty}{\usepackage{upquote}}{}
\ifnum 0\ifxetex 1\fi\ifluatex 1\fi=0 
\usepackage[utf8]{inputenc}
\else 
\usepackage{fontspec}
\ifxetex
\usepackage{xltxtra,xunicode}
\fi
\defaultfontfeatures{Mapping=tex-text,Scale=MatchLowercase}

\fi
\IfFileExists{microtype.sty}{\usepackage{microtype}}{}
\usepackage{graphicx}
\makeatletter
\def\maxwidth{\ifdim\Gin@nat@width>\linewidth\linewidth
\else\Gin@nat@width\fi}
\makeatother
\let\Oldincludegraphics\includegraphics
\renewcommand{\includegraphics}[1]{\Oldincludegraphics[width=\maxwidth]{#1}}
\ifxetex
\usepackage[setpagesize=false, 
unicode=false, 
xetex]{hyperref}
\else
\usepackage[unicode=true]{hyperref}
\fi
\hypersetup{breaklinks=true,
bookmarks=true,
pdfauthor={},
pdftitle={Identifying emerging influential nodes in evolving networks : Exploiting strength of weak nodes},
colorlinks=true,
urlcolor=blue,
linkcolor=magenta,
pdfborder={0 0 0}}
\usepackage[nomarkers]{endfloat}

\begin{document}
\begin{frontmatter}

\title{Identifying emerging influential nodes in evolving networks: Exploiting strength of weak nodes}
\author[{Web Science Center UESTC,Big Data mining and application CIGIT CAS}]{Khushnood Abbas}
\ead{201414060110@std.uestc.edu.cn} 
\author[Big Data mining and application CIGIT CAS]{Mingsheng Shang\corref{c1}}
\ead{msshang@cigit.ac.cn} 
\cortext[c1]{Corresponding Author}
\author[{Big Data Research Center UESTC,Web Science Center UESTC}] {Cai Shi-Min}
\author[Big Data mining and application CIGIT CAS]{Xiaoyu Shi}
\address[Web Science Center UESTC]{ Web Sciences Center, School of Computer Science and Engineering, University of Electronic Science and Technology of China, Chengdu,Sichuan, 611731, PR China}
\address[Big Data Research Center UESTC]{Big Data Research Center, University of Electronic Science and
Technology of China, Chengdu 611731, PR China}
\address[Big Data mining and application CIGIT CAS]{ Chongqing Institute of green and
Intelligent Technology of China, Chinese Academy of Sciences, Chongqing, 401120, P.R. China}
\begin{abstract}
Identifying emerging influential or popular node/item in future on network is a current interest of the researchers. Most of previous works focus on identifying leaders in time evolving networks on the basis of network structure or node's activity separate way. In this paper, we have proposed a hybrid model which considers both, node's structural centrality and recent activity of nodes together. We consider that the node is active when it is receiving more links in a given recent time window, rather than in the whole past life of the node. Furthermore our model is flexible to implement structural rank such as PageRank and webpage click information as activity of the node. For testing the performance of our model, we adopt the PageRank algorithm and linear preferential attachment based model as the baseline methods. Experiments on three real data sets (i.e Movielens, Netflix and Facebook wall post data set), we found that our model shows better performance in terms of finding the emerging influential nodes that were not popular in past.
\end{abstract}
\end{frontmatter}

\section{1. Introduction}\label{introduction}

The influence analysis problem was initiated since the introduction of
the viral marketing phenomena {[}42{]}. Finding influenceres is
considered as many body problem in which topological features play a very
important role {[}4{]}. In general topologically centralized nodes plays
an important role as influencer {[}2{]}, {[}40{]} but influential
phenomena is not always depends on structural centrality of the node.
Sometimes weakly connected nodes can be influential node. Since
considering the collective influence effect top influencers are low
degree nodes as compare to hubs in the network {[}38{]}. Activity of the
node plays a very important role in influential processes, i.e., if the
nodes are hub but not active, it cannot help in spreading processes.
Activity of the node can be quantified as many ways such as rate of in
degree or out degree formation and so on. Here we have considered in
degree formation with temporal effect. Activity of the node can also be
considered how fast the node is making links with other nodes such as in
case of social network how actively user is involved in commenting,
liking and creating posts. Activity of the node can also be considered
as how many friends or followers the user is able to make in recent
past. To contain this information we have made network that contains
both characteristics; such as if user A comment or like on user B's
posts then there is a directed link from A to B. In the same manner if
user C followed user A then there is also a link from C to A, although
our data set lacks this full information. So we have created only
activity based network. If we are able to identify the critical node
having specific characteristics in dynamically evolving network, it will
help us in solving many problems from our daily life such as epidemics
outbreak {[}39{]}, advertisement {[}26{]}, precluding failure of grids or
internet{[}1{]}, managing transport problem {[}50{]} , telecommunication
network and so on. Preferential attachment explains effectively growth
of time evolving real time system. Most recently researchers have found
that the popularity of online contents like news , blog posts , videos,
mobile app download {[}22{]} in online discussion forums and product
reviews exhibits temporal dynamics.{[}52{]}, {[}27{]} has also discussed
how content popularity fluctuate with time. Some items grow ones while
some items can again gain popularity after decay {[}8{]}. In these
cases when items can gain popularity multiple times considering recent
popularity will give benefit. In other way in cases when content gain
popularity many times the whole item or node degree will not be a good
predictor, although recent popularity gain may help as it is found in predicting
future popularity gain {[}37{]}, {[}10{]}, {[}41{]}, {[}37{]}. In some network such
as paper citation the rate of attracting new node decays due to aging
factor {[}41{]}, {[}37{]}. Therefore we can say for making prediction
aging factor also matters. Although in some network such as citation
network the new link formation will also matter on the fact that this
node has been linked with which node. In other words if a paper has been
cited by a popular scientist then it's probability increases that other
will cite or make link with this node.

Problem of finding influential user or node in network is being studied
for quite long time such as {[}28{]}, {[}13{]}, {[}38{]}. They have
considered static network and did not taken into consideration temporal
dynamics of the node. Since we know by time how network structure are
changing there for its node centrality also get affected so to consider
only static view of the network is not good in knowledge discovery. Node
centrality may vary from network to network especially when we consider
the dynamics of the node. Some researchers have solved this problem
using supervised classification problem such as {[}45{]}. Authors in reference {[}5{]} have
proposed Popularity based predictors(PBP) for predicting popular item or
node using bipartite network. It is ``preferential attachment'' theory
based model; they show that instead of whole degree of the node recent
degree gain is a good predictor. In addition they also proposed the
novel entries in the top list those were not in past time window.
Further ({[}22{]}) also came up that recent degree of the node is a good
predictor for future popularity of the node. Some researchers have found
the aging or decay factor also in the time evolving node's degree
{[}41{]}, {[}51{]}, {[}15{]}. {[}12{]}, {[}30{]}. PageRank based metrics
has also been exploited by many researchers. The authors in {[}29{]}
applied PageRank algorithm on twitter conversational network, they have
considered metrics such as acceleration of the graph also. Similarly
authors in {[}36{]} have proposed userRank based on the tweet relevance
for user influence. {[}16{]} came up diversity-dependent influence score
(DIS) in tweeter conversational graph for user influence by exploiting
the social diversity and influence propagation phenomena. It has also
employed the PageRank philosophy. The authors in reference {[}14{]} have
proposed ``Influence Rank'' based recursive concept of user influence.
Authors in {[}33{]} have proposed tweet correlation based financial
users prediction, they supposed if user A does similar tweet than user B
then they will influence each other. Although this method will not apply in macro level analysis, such as if we are interested in viral node. {[}21{]} proposed three
algorithm based on PageRank namely ``InfRank'', ``LeadRank'' and
``DiscussRank''. InfRank is the ability of a user to spread information
and to be re tweeted by other influential users. LeadRank quantifies the
ability of user to stimulate other user to re tweet and mention from
other user. DiscussRank identifies the active users who initiate
discussion and involved others in discussion by replying and initiating
tweets. {[}7{]}, {[}31{]} has applied PageRank algorithm on citation
data sets, they have found Google PageRank and citation positively
correlated. {[}48{]} has proposed `TwitterRank' algorithm extension of
PageRank algorithm for identifying topic based influential users in
micro blogging network such as twitter.{[}43{]} has considered recent
citation with Google PageRank to predict the future citation gain of the
paper. A recent work by {[}46{]} has solved this problem by considering
information flow pattern in the network. Authors in reference {[}47{]}
have solved this problem considering static network and exploiting
k-core decomposition method.

\section{2. Preliminaries}\label{preliminaries}

\subsection{2.1 Problem definition}\label{problem-definition}

In this script we have given a model to find top-k emerging influential
nodes in evolving network. We have modeled the problem by taking
snapshot of the network at different time. Snap shot contains all the
information of the link before time \(t\) only. The prediction problem
is, considering node \(n\) in any system at time \(t\) and past time
window \(T_P\), predicting the ranking of the node after future time
window \(T_F\) on the basis of number of links it has received in past
time window \(T_P\). So in case of user item bipartite network {[}31{]},
{[}44{]}, {[}17{]} we can consider how many user have bought/rated as
receiving link of the node \(n\). In case of other directed network, how
many link node \(n\) has received. So the prediction problem can be
defined as given snap shot of the network at time \(t\) and past time
window \(T_P\) we have to predict node's ranking in future at \(t+T_F\)
time on the basis of links received during \((t,T_F)\), where \(T_F\) is
future time window. Second we have to predict the novel entries or
emerging influential node that were not in top \(N\) list before time
\(t-T_P\).

\subsection{2.2 Baseline method}\label{baseline-method}

\subsection{PageRank}\label{PageRank}

PageRank given by {[}6{]} was developed to rank webpages on internet
for Google search engine optimization purpose. It can be applied in
other networks also where structural property of the node plays an
important role such as information diffusion {[}48{]}, scientific
paper/author ranking {[}43{]} etc. PageRank has also been applied on
bipartite network {[}3{]}, {[}19{]} PageRank algorithm can be given as
follows:- If site \(n_i\) have link to site \(n_j\) there will be a
directed link between node \(n_i\) and node \(n_j\) (\(n_i->n_j\)). If
page/node \(n_j\) has \(S_i\) set of link to other pages/nodes then page
will distribute its importance in \(|l_j|\)(number of nodes in set
\(|S_i|\)) nodes equally. The transition matrix of a graph \(A\) can be
given as follows-

\[
\begin{aligned}
A_{ij} = \left\{ {\begin{array}{*{20}c}
{1/l_j {\ }if{\ }n_j \in s_i } \\
{0{ \quad } Otherwise} \\
\end{array}} \right\}
\end{aligned}
\]

Since there can be pages that do not have link to other pages although
they are being pointed by other page, also known as \emph{dangling
nodes} , so new transformed matrix can be given as-

\[
\begin{aligned}
{\rm S = A + N}_{{\rm cd}} 
\end{aligned}
\]

where \(N_{\rm cd}\) matrix have all the elements zero except for
\textbf{dangling nodes'} column which are \(1/N\) where \(N\) is the
number of rows or nodes in matrix. Easy to find that those columns are
normalized that sums to one for making column stochastic matrix. Now the
PageRank of \emph{dangling nodes} will not be zero. Since random surfer
will follow the link from one page to another, suppose that random
surfer follows the PageRank (follows S) with probability \(({\alpha})\)
then there is (\(1-\alpha\)) probability that he will choose a random
page. So now PageRank matrix also known as Google matrix \(M\) can be
given as-

\[
\begin{aligned}
M = \alpha S + \frac{{(1 - \alpha )}}{n}{I_n}
\end{aligned}
\]

where \(I_n\) is matrix of size \(n*n\), it's every element as one .
Since \textbf{Google Matrix} \(M\) is combination of stochastic matrix
and all the entries are positive which implies that M is primitive and
irreducible. So PageRank vector \(PR\) can be calculated using power
method as \(PR^k = M.PR^{^{(k - 1)} }\) it will eventually converge to a
static vector which is PageRank .

\section{3. Model}\label{model}

Importance of node depends on structural position as well as activity
(e.g rate of link formation) of the node. If node is not active in
current recent time it's structural centrality have no value in some
context such as information spread. A webpage might be ranked high due to
other important webpages are pointing to it but if people are not
clicking on the link after recommendation it means it is not an
important page at least in temporal sense. So to develop a good model
for predicting influential node we have considered both characteristics
in our model.

\begin{equation}
s_n (t,T_P ) \propto s_c (n,t)*d_c (n,t,T_p )
\end{equation}

Where \(s_n (t,T_P )\) predicted rating score of node \(n\) at time
\(t\) given past time window \(T_p\), \(s_c (n,t)\) structural
centrality metric of node such as degree, closeness, PageRank at time
\(t\) etc and \(d_c (n,t,T_p )\) is dynamic centrality of the node at
time \(t\) given some past time window \(T_p\) such as {[}51{]},
{[}5{]}, {[}32{]} has given model considering node's
activity. \(d_c (n,t,T_p )\) can vary according to need such as in
webpage ranking structural property can be PageRank while dynamic
centrality can be a score based on it's click over time. PageRank is one
of the best model that considers structural feature of the network so we
have considered PageRank as structural centrality metric. The PageRank
code is given by {[}34{]}. The original PageRank algorithm was given by
{[}6{]}. The dynamic metric we have considered given by {[}51{]}.
Although the generic model above can be implemented according to need of
the problem such as if we care about a critical node to prevent the node
damage, we can use Betweenness centrality as a structural centrality and
dynamic centrality as it's `work done' such as in case of Internet
router network and power grid network. Here we are considering PageRank
as structural metric since it's wide application in different areas such
as academic articles and authors {[}7{]}, {[}11{]}, {[}49{]} , image
ranking {[}24{]}, urban road ranking {[}23{]}, protein interaction
network {[}20{]}, software ranking on the basis of Procedure Call
Network (PCN), since large software contains many procedures that call
to each other {[}9{]} So our one of specific model is as following we name it m1.

\begin{equation}
s_n (t,T_P ) =K \sum\limits_{n',t_n > = T_p } {PR_n (t) * {\exp (\gamma (t_n - t))} } 
\end{equation}

Where K is a normalization constant which normalizes the score of all node which sums to 1. Here \(t_n\) is the time when node \(n\) received link, \(T_P\) is the
past time window. \(\gamma\) is decay rate, \(PR_n (t)\) is PageRank
score of the node \(n\), n' is the set of rest of the node from which n
has received link. \(t_n > = T_p\) condition tells that we consider only
link that is formed after \(t-T_P\) time. Considering only recent
activity (\(T_P\)) of the node saves computation cost since in real time
network size is very large. We are considering in degree of the node
one can consider out degree also if activity of the node is based on out degree.

\subsection{When there are only few active nodes in a
system}\label{when-there-are-only-few-active-nodes-in-a-system}

Since from the above model if the node is not receiving any link after
time (\(t-T_p\)), its score will be zero. To overcome this weakness we
assume even if node has not received any link during past time window
\(T_P\) there is a probability of getting new link in future according
to its structural centrality. Suppose probability of a node for getting
in-links depends on recent gain in new links then we can say probability
(\(P_{n/t,T_p }\)) given past time window \(T_P\)) of node \(n\) at time
\(t\) of getting links in future is -

\begin{equation}
P_{n/t,T_p } \propto \sum\limits_{n',t_n < = T_P } {\exp (\gamma (t_n - t))} 
\end{equation}

So predicted rating score \(s_n (t,T_P )\) of a node can be given as
follows-

\begin{equation}
s_n (t,T_P ) =L * PR_n (t)(1 + P_{n/t,T_p } )
\end{equation}

Where L is a normalization constant which normalizes the score of all node which sums to 1. The above predictor makes sure that even if node has not received any
link in recent past it will be ranked according to its structural
centrality. Although nodes which are more structurally central, and also
active in recent past will be given more score. In addition if node's
structural centrality is not high but recently it is very active then
score will also be high which is why we call it ``strength of a
weak node''. Such as in case of web page ranking if we want to consider
both the centrality.

\subsection{When system contains, recent activity as well as total
popularity
followers.}\label{when-system-contains-recent-activity-as-well-as-total-popularity-followers.}

If in any system people follows rich get richer phenomena as well as
recent behavior. Then we can model as follows, suppose with
probability \(\delta\) they follow already popular node or item and with
probability \((1 - \delta )\) they will follow recent behaviors of
their peers. Such as in case of author citation network people do care
about scientists' popularity and also if any new scientist does a
potential discovery. In a system where people follow recent behaviors
of their peers the model in equation (4) will be a good predictor for
future popularity gain of the node. Although it can be modeled as recent
degree gain also, since if people follow recent behavior then recent
degree gain ``must'' be a good predictor for future popularity or link
gain of the node. We have considered decay effect because nodes in most
of the time evolving systems show competition for getting links which
causes aging effect in other nodes. Introduction of aging phenomena
helps in identifying emerging influential nodes.

\begin{equation}
s_n (t,T_P ) = M  (\delta PR_n (t) + (1 - \delta )P_{n/t,T_p } )
\end{equation}

Where M is a normalization constant which normalizes the score of all node which sums to 1.
\section{4. Experiment and Results}\label{experiment-and-results}

\subsection{About Dataset}\label{about-dataset}

We have used Netflix data set the original Netflix data has \(480189\)
users, \(100480507\) ratings and \(17770\) objects. The original
Facebook data set have \(42390\) , \(39986\) objects and \(876993\)
links. While data preparation we have sampled small subset from each by
randomly choosing users who have rated at least \(20\) movies. The
original rating was in the form of numerical \(1-5\), we have considered
the link between the user and object which object have received higher
than two ratings. For all the three data sets Facebook, Movielens and
Netflix the time is considered in days. In case of Facebook if user has
posted/like or commented on other user's wall there will be a link
between the user and the wall. The link between user and its own wall
post have not been considered to avoid self-influence. The data
description after cleaning process -

\begin{itemize}
\item
\textbf{Netflix} data contains \(4960\) users,\(16599\) movies and
\(1249058\) links, data was collected during(1st Jan \(2000\)
--\(31st\) Dec 2005).
\item
\textbf{MovieLens} data set contains \(7533\) movies, \(864581\) links
and \(5000\) users and data was collected during(\(1st\) Jan \(2002\)
--\(1st\) Jan 2005).
\item
\textbf{Facebook} data contains \(40981\) set of users and their
\(38143\) wall post activity and \(855542\) links, during period of
(14 Sep \(2004\)--\(22nd\) Jan \(2009\)).
\end{itemize}

\subsection{Evaluation metrics}\label{evaluation-metrics}

Following evaluation metrics are adopted to measure the accuracy of the
proposed model including
\emph{precision}\((P_n)\),\emph{novelty}\((Q_n)\), \emph{Area Under
Recieving Operating Characteristic}(\(AUC\)) and Kendall's
Tau(\(\tau\)).

\begin{itemize}
\tightlist
\item
\emph{Precision} is defined as the fraction of objects that are
predicted also lie in the top \(N\) object of true ranking {[}25{]}.
\end{itemize}

\[
\begin{aligned}
{p_n = \frac{{D_n }}{n}}
\end{aligned}
\] Where \(D_n\) is the number of common objects between predicted and
real ranking. \(n\) is the size of list to be ranked. It's value ranges
in {[}0,1{]}, higher value of \((P_n)\) is better.

\begin{itemize}
\tightlist
\item
\emph{Novelty(\(Q_n\))} is a metric to measure the ability of a
predictor to rank the items in top \(n\) position that was not in top
\(n\) position in previous time window. We call these new entries as
``potential items'' or emerging leaders throughout the script. If we
denote the predicted object as (\(P_po\)) and potential true object as
\(P_ro\) , then the novelty of a model is given by {[}5{]}-
\end{itemize}

\[
\begin{aligned}
Q_n = P_{po} /P_{ro}
\end{aligned}
\]

\begin{itemize}
\tightlist
\item
\emph{AUC} measures the relative position of the predicted item and
true ranked items. Suppose predicted item list is (\(L_pn\)) and real
item list is (\(L_rn\)). if \(s_op \in L_{pn}\) and
\(s_rp \in L_{rn}\) is score of object in predicted then \emph{AUC} is
given by -
\end{itemize}

\[
\begin{aligned}
AUC = \frac{{\sum\limits_{op \in L_{pn} } {\sum\limits_{rp \in L_{rn} } {I(s_{pn} ,s_{rn} )} } }}{{\left| {L_{pn} } \right|\left| {L_{rn} } \right|}}
\end{aligned}
\]

where, \[
\begin{aligned}
I(s_{pn} ,s_{rn} ) = \left\{ 
{\begin{array}{*{20}c}
{0 \Leftarrow s_{pn} < s_{rn} } \\
{0.5 \Leftarrow s_{pn} = s_{rn} } \\
{1 \Leftarrow s_{pn} > s_{rn} } \\
\end{array}} \right.
\end{aligned}
\]

\begin{itemize}
\tightlist
\item
\emph{Kendal's Tau(\(\tau\))} measures the correlation between
predicted and actual ratings. It varies between \(-1\) and \(+1\).
\(\tau =1\) when predicted and actual are identical,\(\tau =0\) when
both ranking is independent and \(\tau =-1\) shows they perfectly
disagree. It can be given as-
\end{itemize}

\[
\begin{aligned}
{\tau}= \frac{{C - D}}{{C + D}}
\end{aligned}
\]

Where \(C\) is the number of concordant pairs and \(D\) is the number of
discordant pairs.

\subsection{Results and discussion}\label{results-and-discussion}

\subsection{Popularity-based
predictor}\label{popularity-based-predictor}

The authors in {[}5{]} came up with Popularity based predictor(PBP). It exploit the
\emph{preferential attachment} theory , which states that popularity
increases cumulatively; the rate of new link (Either item receives
rating in case of Movielens, or a friend like or comments in case of
Facebook wall post activity) formation for any node is proportional to
the observed number of links which node has received in past. If an item
is popular at time \({\rm t}\), then it will probably become popular due
to the condition that current degree of an item
\({\rm k}_o {\rm (t} {\rm )}\) is a good predictor of its future
popularity. Further ({[}22{]}, {[}5{]}) have found that current degree
is a good predictor of items' future popularity.{[}5{]} proposes to
calculate the prediction score of an item at time \(t\) can be given as
follows-

\begin{equation}
{\rm s}_o {\rm (t} {\rm ,T}_{\rm p} {\rm ) = k}_o {\rm (t} ) - \lambda {\rm k}_o {\rm (t} {\rm , T}_{\rm P} {\rm )}
\end{equation}

Where \(\Delta {\rm k}_o {\rm (t} {\rm , T_P)}\) is the rating/links
received in past time window \(T_P\) from \(t\).
\(\lambda \in {\rm [0, 1] }\) , note that \(\lambda = 0{\rm }\) gives
the total popularity and for \(\lambda = 1{\rm }\) it gives recent
popularity. Throughout the script by popularity we mean number of
ratings or links received by item or node.

To check the accuracy and robustness of our proposed model we have
considered various data sets such as MovieLens, Netflix and Facebook
wall post activity. Our method out performs specially in predicting
emerging entries. While performing experiments we have selected 10
random time and constructed network on the basis of prior link
formation. From randomly chosen point we measure the past time window
\(T_P\) and future time window \(T_F\). We have constructed the network
up-to the randomly selected time and calculated PageRank score of the
nodes. In case of our proposed hybrid PageRank we have considered how
many ratings the node has received in the past time window. We took
average of 10 results to make sure robustness of our model.

\subsubsection{\texorpdfstring{Relationship between decay rate
(\(\gamma\)) and teleportation parameter 
(\(\alpha\))}{Relationship between decay rate (\textbackslash{}gamma) and teleprtation paramter (\textbackslash{}alpha)}}\label{relationship-between-decay-rate-gamma-and-teleprtation-paramter-alpha}

Below are the empirical results on relationship between PageRank
teleportation parameter (\(\alpha\)) and decay rate (\(\gamma\)). 
We have considered different values of both
parameters to find that best fit with the data. In all the data sets
we have found the optimal value of \(\gamma\) and \(\alpha\) is
\(\tilde 0.1\). Decay rate more affect novelty prediction \(\gamma\).
Although the optimal value is around \(0.1\) which maximizes the
precision for these data sets. In manual parameter setting one can
choose higher value of \(\gamma\) to recommend more novel items.

\includegraphics{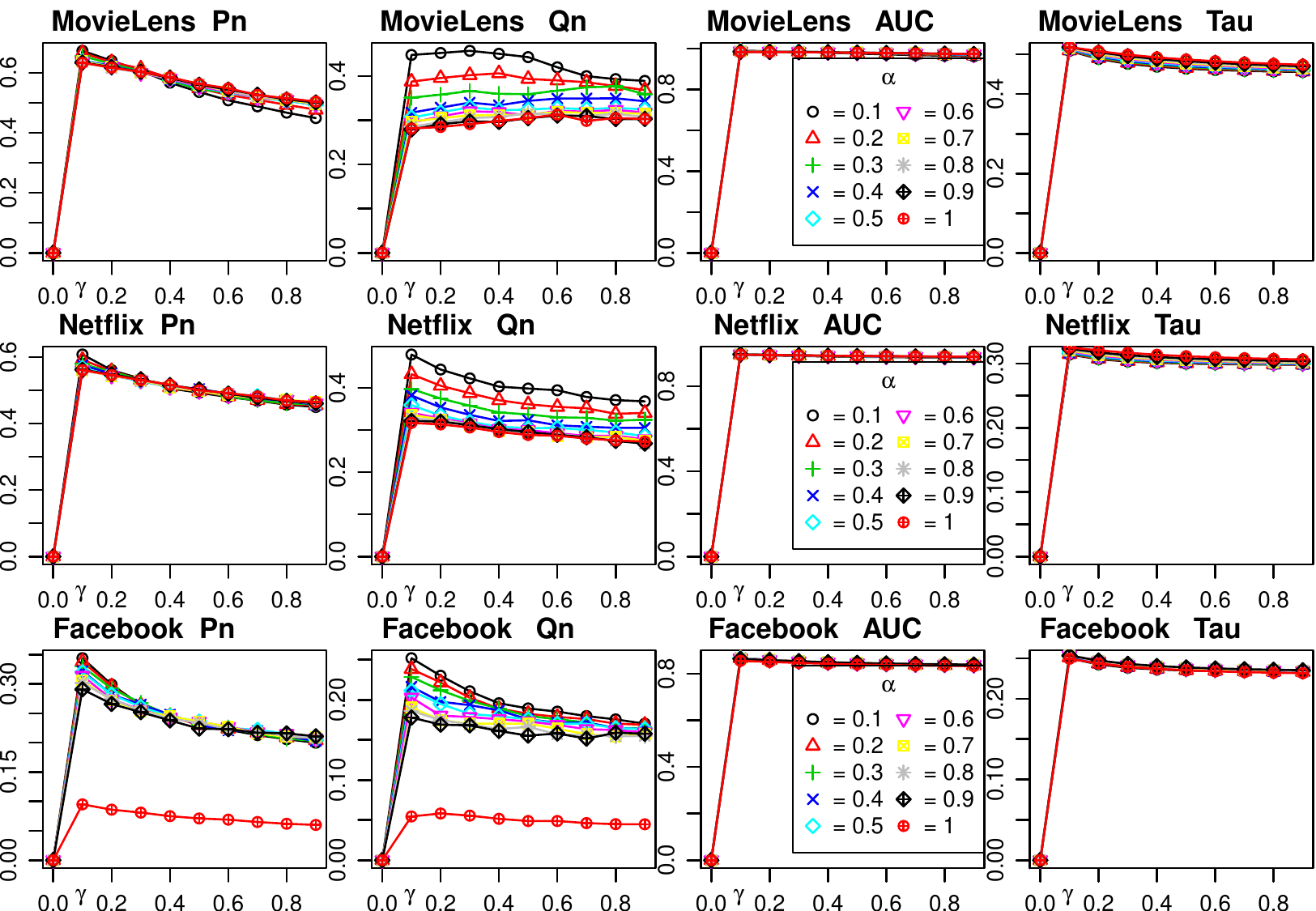}

\begin{quote}
Figure 1: Effect of both the parameters on our model and, on accuracy.
Here we have consider \(T_P\) and \(T_F\) both as \(30\) days. The Rank
correlation is for whole data while other metrics is for top \(100\)
items.
\end{quote}

In figure {[}Figure 2{]} we have compared the result with baseline
methods. We have considered the future and past time window \(T_F\) ,
\(T_P\) respectively same. We can see our model have better prediction
accuracy for novel or emerging node prediction. Our model has better
\textbf{precision} accuracy with respect to PageRank . \textbf{AUC}
metric also shows better performance over PageRank while with respect to
PBP it is not very low. It shows our predictors robustness since it
shows better performance for top items means it support preferential
attachment theory therefore it can be applied in scale free networks.
It's rank correlation is also considerable means it's performance will
not go down no matter how big the recommendation list size is. From
{[}Figure 2{]} we can see the correlation coefficient is not better than
the other two the reason the other metrics only considers top \(100\)
items while correlation is for whole nodes. We put \(\tau\) here to show
the robustness of our model that it's performance not much affected in
digging novel entries.

\includegraphics{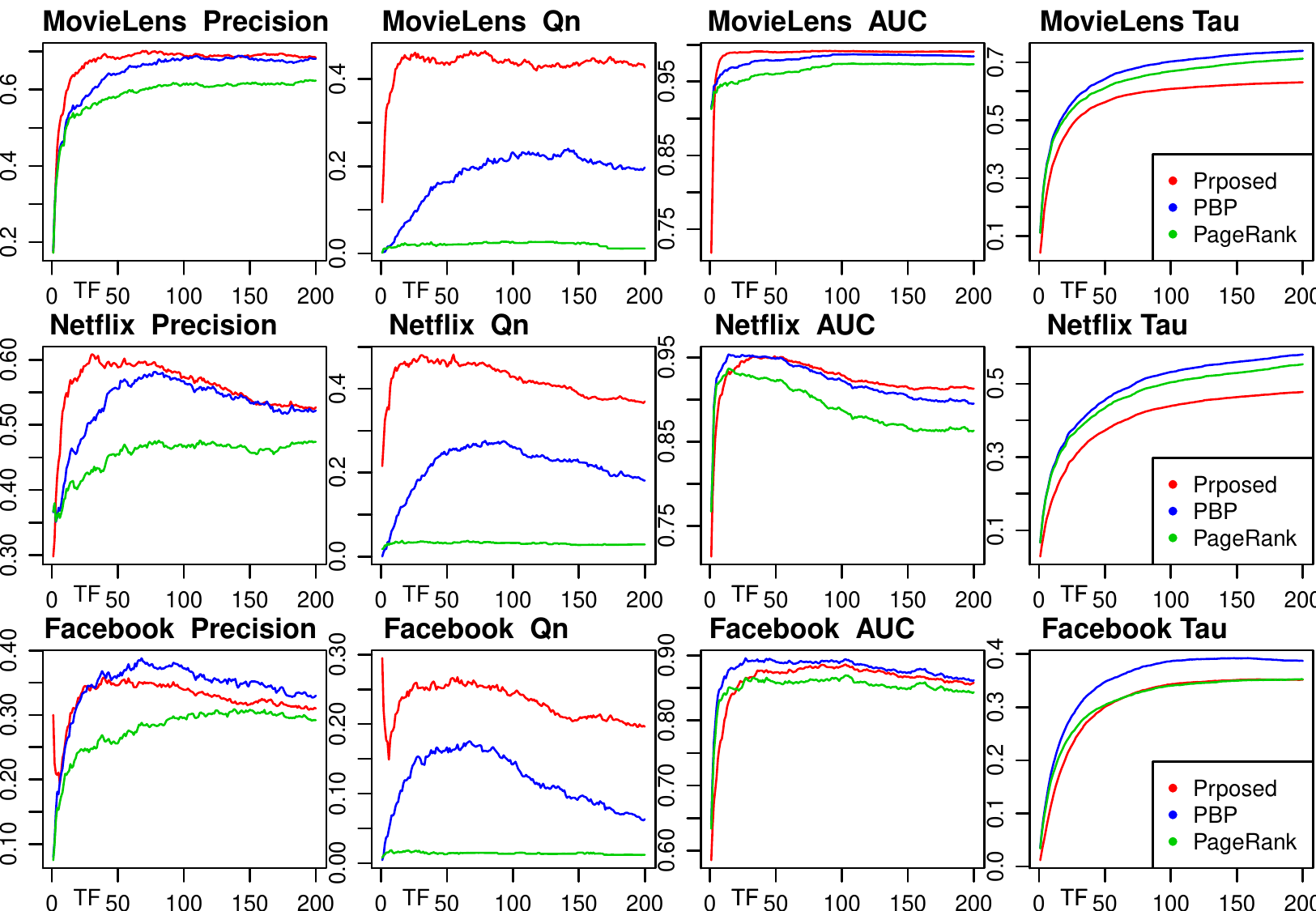}

\begin{quote}
Figure 2: The above figure shows the performance of the predictor for
different values on future time window \(T_F\). Red line shows the
performance of our proposed predictor while blue shows the base
predictor PBP and green line PageRank predictor. \(T_F\) and \(T_P\) are
considered same upto \(200\) days. This is result of model m1.
\end{quote}

\subsection{Results using equation (4)}\label{results-using-equation-4}

The above results are obtained using equation (2). But as we see some
times we need equation(4) to rank the nodes. In {[}Figure 3{]} we have
considered past and future time window as 30 days for all the data sets.
For evaluating Precision (\(P_n\)), Novelty(\(Q_n\)) and \(AUC\) we have
considered top 100 elements. While \(Tau\) is ranking correlation which
works on the whole data.

\includegraphics{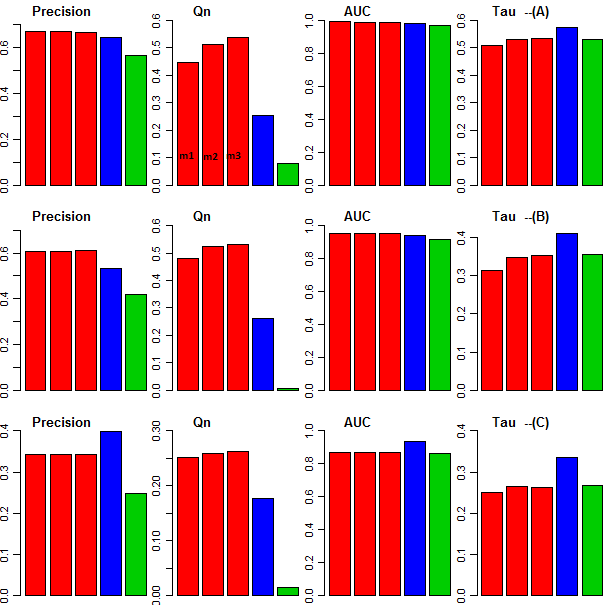}

\begin{quote}
Figure 3: The above figure shows the comparison with different
predictors. The read color bars m1,m2,m3 are for our proposed model in
equation 2,4,5 from left to right respectively. These results are for
different data sets such as MovieLens(A), Netflix(B) and Facebook(C).
The blue color bar is for Populairty Based Predictor and green one is
for PageRank. Past and future time window is considered as 30 days.
\end{quote}

\subsection{Results using equation (5) when people follows popularity as
well as recent activity of the
node.}\label{results-using-equation-5-when-people-follows-popularity-as-well-as-recent-behavior.}

The model given in equation (5) considers both the phenomena. If in some
system people follows popularity as well as recent activity of the node
then this model will work. We have considered the node attracts new link
in a system with probability \(\delta\) according to its structural or topological centrality and with probability (\(1-\delta\)) according to
recent activity of the node. The results on three data sets are as
follows-

\includegraphics{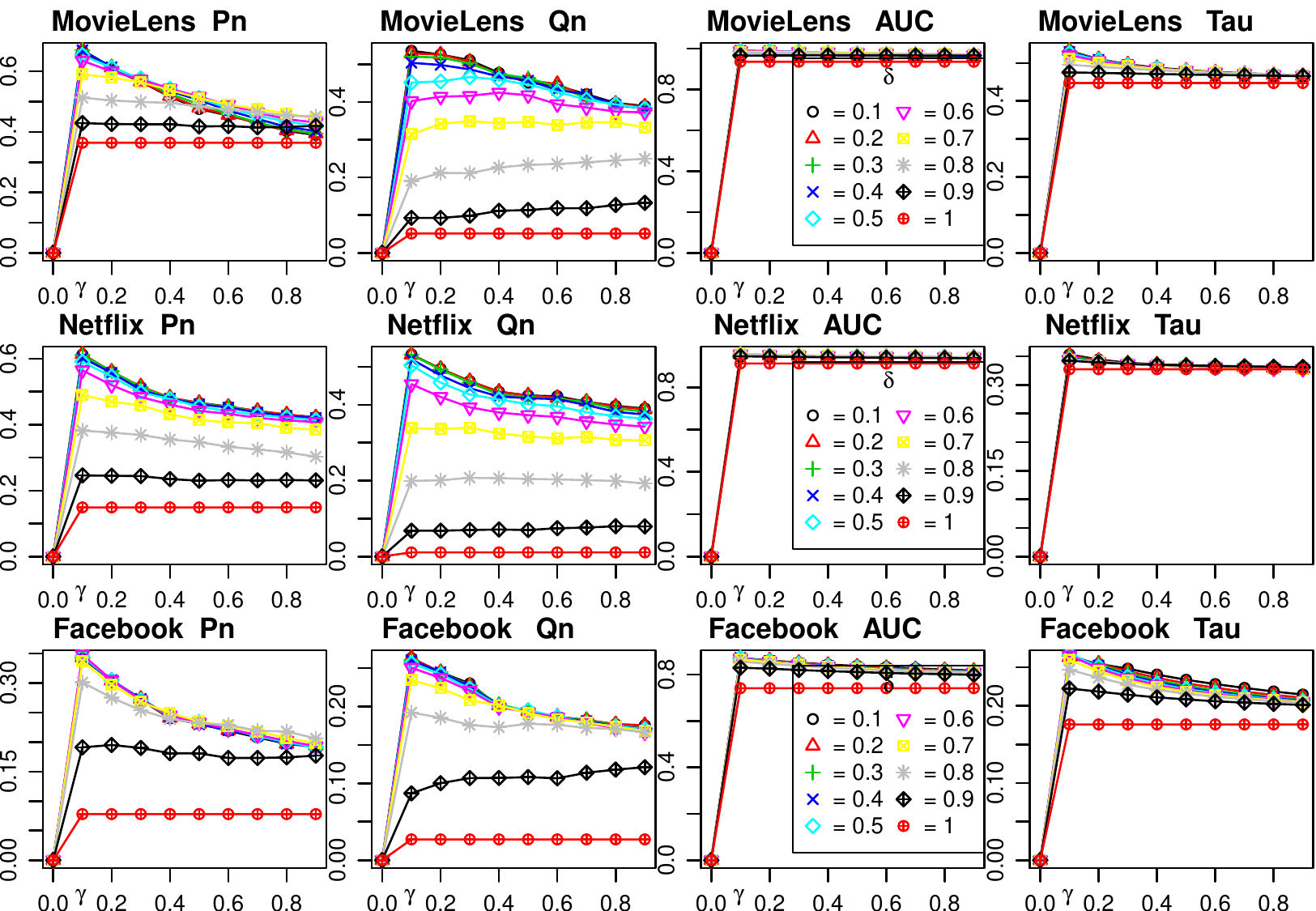}

\begin{quote}
Figure 4: The above figure shows the effect of different parameters on
the results using model given in eq. (5). \(\delta\) is coupling
probability for the two different centrality.
\end{quote}

From the {[}Figure 4{]} we can find that lower value of \(\delta\) gives
better results. Lower value of delta gives higher weight to dynamic
centrality in eq.(5). In other words on these data sets we can say
people follow recent activity of the node.'

\section{5. Conclusion}\label{conclusion}

User attention can be seen as a scarce and valuable resource and its
aggregation as ``attention economy'' {[}18{]}. As user have limited time
and energy therefore they cannot give their full resource to dig
quality items in this universe of online web. It is like searching for
the star without any tool. Although Recommender system helps user to
find items of their interest. But whether Recommender system is able to
identify all the quality items, is a quest. Being a machine judging
quality items is not possible for the Recommender system without any
prior history of the item. Recommending novel item at random can cause
``shock'' to user, which can cost revenue of the enterprise. On the
other hand appropriate novel items can help in getting user attention
therefore revenue generation. As novelty and popularity are two main
causes found for getting user attention {[}18{]}. To overcome the
``novelty-shock'' problem content provider selects many strategies one
of them is promoting their products using social media as a
``word-of-mouth'' or ``viral marketing'' strategy. To implement these
strategy they select the leader or popular user to promote their
product. In this work we have given solution to identifying those
influential node. Although our model can be applied any time evolving
network. Here we have used structural information on the basis of user
activity (Facebook case). The best result will come when we have
topological (friendship network) as well as user activity of the node
such as (wall post, liking and sharing etc). In this paper we have given
model that identifies emerging influential node in time evolving
network. To find the leader node or items on social media, we have
considered both temporally evolving dynamics as well as structural
characteristics of the network. We have found that if we consider recent
temporal effects of the node with PageRank then future popularity
prediction accuracy improves. Our model predicts the emerging
influential nodes, that were not in past popular list without any
significant cost of accuracy. We have considered PageRank algorithm for
ranking nodes on the basis of structural or topological characteristics.
We have also considered the temporal dynamics of the node to consider
recent activity of the node.

\section{Acknowledgement}\label{acknowledgement}

This work is done under NSFC project grant no:61673086.

\section*{References}\label{references}
\addcontentsline{toc}{section}{References}

[1] \hyperdef{}{ref-Albertux5f2004}{\label{ref-Albertux5f2004}}
Albert, R\a'eka, Istv\a'an Albert, and Gary L. Nakarado. 2004.
``Structural Vulnerability of the North American Power Grid.''
\emph{Physical Review E} 69 (2). American Physical Society (APS).
\href{http://doi.org/10.1103/physreve.69.025103}{doi:10.1103/physreve.69.025103}.

[2] \hyperdef{}{ref-Albertux5f2000}{\label{ref-Albertux5f2000}}
Albert, R\a'eka, Hawoong Jeong, and Albert-L Barabasi. 2000. ``Error and
Attack Tolerance of Complex Networks.'' \emph{Nature} 406 (6794). Nature
Publishing Group: 378--82.
\href{http://doi.org/10.1038/35019019}{doi:10.1038/35019019}.

[3] \hyperdef{}{ref-Allesinaux5f2009}{\label{ref-Allesinaux5f2009}}
Allesina, Stefano, and Mercedes Pascual. 2009. ``Googling Food Webs: Can
an Eigenvector Measure Species Importance for Coextinctions?'' Edited by
Philip E. Bourne. \emph{PLoS Comput Biol} 5 (9). Public Library of
Science (PLoS): e1000494.
\href{http://doi.org/10.1371/journal.pcbi.1000494}{doi:10.1371/journal.pcbi.1000494}.

[4] \hyperdef{}{ref-Altarelliux5f2013}{\label{ref-Altarelliux5f2013}}
Altarelli, F, A Braunstein, L Dall'Asta, and R Zecchina. 2013.
``Optimizing Spread Dynamics on Graphs by Message Passing.'' \emph{J.
Stat. Mech.} 2013 (09). IOP Publishing: P09011.
\href{http://doi.org/10.1088/1742-5468/2013/09/p09011}{doi:10.1088/1742-5468/2013/09/p09011}.

[5] \hyperdef{}{ref-DBLP:journalsux2fadvcsux2fZengGMZ13}{\label{ref-DBLP:journalsux2fadvcsux2fZengGMZ13}}
An Zeng, Stanislao Gualdi, Matus Medo, and Yi-Cheng Zhang. 2013. ``Trend
Prediction in Temporal Bipartite Networks: The Case of MovieLens,
Netflix, and Digg.'' \emph{Advances in Complex Systems} 16 (4-5).
\href{http://doi.org/10.1142/S0219525913500240}{doi:10.1142/S0219525913500240}.

[6] \hyperdef{}{ref-Brinux5f2012}{\label{ref-Brinux5f2012}}
Brin, Sergey, and Lawrence Page. 2012. ``Reprint of: The Anatomy of a
Large-Scale Hypertextual Web Search Engine.'' \emph{Computer Networks}
56 (18). Elsevier BV: 3825--33.
\href{http://doi.org/10.1016/j.comnet.2012.10.007}{doi:10.1016/j.comnet.2012.10.007}.

[7] \hyperdef{}{ref-Chenux5f2007}{\label{ref-Chenux5f2007}}
Chen, P., H. Xie, S. Maslov, and S. Redner. 2007. ``Finding Scientific
Gems with Google's PageRank Algorithm.'' \emph{Journal of Informetrics}
1 (1). Elsevier BV: 8--15.
\href{http://doi.org/10.1016/j.joi.2006.06.001}{doi:10.1016/j.joi.2006.06.001}.

[8] \hyperdef{}{ref-cheng2016cascades}{\label{ref-cheng2016cascades}}
Cheng, Justin, Lada A Adamic, Jon M Kleinberg, and Jure Leskovec. 2016.
``Do Cascades Recur?'' In \emph{Proceedings of the 25th International
Conference on World Wide Web}, 671--81. International World Wide Web
Conferences Steering Committee.
\href{http://doi.org/10.1145/2872427.2882993}{doi:10.1145/2872427.2882993}.

[9] \hyperdef{}{ref-DBLP:journalsux2fcorrux2fabs-1003-5455}{\label{ref-DBLP:journalsux2fcorrux2fabs-1003-5455}}
Chepelianskii, A. D. 2010. ``Towards Physical Laws for Software
Architecture.'' \emph{CoRR} abs/1003.5455.
\url{http://arxiv.org/abs/1003.5455}.

[10] \hyperdef{}{ref-Craneux5f2008}{\label{ref-Craneux5f2008}}
Crane, R., and D. Sornette. 2008. ``Robust Dynamic Classes Revealed by
Measuring the Response Function of a Social System.'' \emph{Proceedings
of the National Academy of Sciences} 105 (41). Proceedings of the
National Academy of Sciences: 15649--53.
\href{http://doi.org/10.1073/pnas.0803685105}{doi:10.1073/pnas.0803685105}.

[11] \hyperdef{}{ref-Dingux5f2009}{\label{ref-Dingux5f2009}}
Ding, Ying, Erjia Yan, Arthur Frazho, and James Caverlee. 2009.
``PageRank for Ranking Authors in Co-Citation Networks.'' \emph{J. Am.
Soc. Inf. Sci.} 60 (11). Wiley-Blackwell: 2229--43.
\href{http://doi.org/10.1002/asi.21171}{doi:10.1002/asi.21171}.

[12] \hyperdef{}{ref-Ferraraux5f2014}{\label{ref-Ferraraux5f2014}}
Ferrara, Emilio, Roberto Interdonato, and Andrea Tagarelli. 2014.
``Online Popularity and Topical Interests Through the Lens of
Instagram.'' In \emph{Proceedings of the 25th ACM Conference on
Hypertext and Social Media - HT 14}. Association for Computing Machinery
(ACM).
\href{http://doi.org/10.1145/2631775.2631808}{doi:10.1145/2631775.2631808}.

[13] \hyperdef{}{ref-Goyalux5f2011}{\label{ref-Goyalux5f2011}}
Goyal, Amit, Wei Lu, and Laks V.S. Lakshmanan. 2011. ``SIMPATH: An
Efficient Algorithm for Influence Maximization Under the Linear
Threshold Model.'' In \emph{2011 IEEE 11th International Conference on
Data Mining}. Institute of Electrical \& Electronics Engineers (IEEE).
\href{http://doi.org/10.1109/icdm.2011.132}{doi:10.1109/icdm.2011.132}.

[14] \hyperdef{}{ref-DBLP:confux2fsocialcomux2fHajianW11}{\label{ref-DBLP:confux2fsocialcomux2fHajianW11}}
Hajian, Behnam, and Tony White. 2011. ``Modelling Influence in a Social
Network: Metrics and Evaluation.'' In \emph{PASSAT/SocialCom 2011,
Privacy, Security, Risk and Trust (PASSAT), 2011 IEEE Third
International Conference on and 2011 IEEE Third International Conference
on Social Computing (SocialCom), Boston, MA, USA, 9-11 Oct., 2011},
497--500. IEEE.
\href{http://doi.org/10.1109/PASSAT/SocialCom.2011.118}{doi:10.1109/PASSAT/SocialCom.2011.118}.

[15] \hyperdef{}{ref-Zhuux5f2003}{\label{ref-Zhuux5f2003}}
Han Zhu, Xinran Wang, and Jian-Yang Zhu. 2003. ``Effect of Aging on
Network Structure.'' \emph{Physical Review E} 68 (5). American Physical
Society (APS).
\href{http://doi.org/10.1103/physreve.68.056121}{doi:10.1103/physreve.68.056121}.

[16] \hyperdef{}{ref-DBLP:confux2fairsux2fHuangLLC13}{\label{ref-DBLP:confux2fairsux2fHuangLLC13}}
Huang, Pei-Ying, Hsin-Yu Liu, Chun-Ting Lin, and Pu-Jen Cheng. 2013. ``A
Diversity-Dependent Measure for Discovering Influencers in Social
Networks.'' In \emph{Information Retrieval Technology - 9th Asia
Information Retrieval Societies Conference, AIRS 2013, Singapore,
December 9-11, 2013. Proceedings}, edited by Rafael E. Banchs, Fabrizio
Silvestri, Tie-Yan Liu, Min Zhang, Sheng Gao, and Jun Lang,
8281:368--79. Lecture Notes in Computer Science. Springer.
\href{http://doi.org/10.1007/978-3-642-45068-6_32}{doi:10.1007/978-3-642-45068-6\_32}.

[17] \hyperdef{}{ref-Huangux5f2007}{\label{ref-Huangux5f2007}}
Huang, Zan, Daniel D. Zeng, and Hsinchun Chen. 2007. ``Analyzing
Consumer-Product Graphs: Empirical Findings and Applications in
Recommender Systems.'' \emph{Management Science} 53 (7). Institute for
Operations Research; the Management Sciences (INFORMS): 1146--64.
\href{http://doi.org/10.1287/mnsc.1060.0619}{doi:10.1287/mnsc.1060.0619}.

[18] \hyperdef{}{ref-Hubermanux5f2012}{\label{ref-Hubermanux5f2012}}
Huberman, Bernardo A. 2012. ``Social Computing and the Attention
Economy.'' \emph{Journal of Statistical Physics} 151 (1-2). Springer
Science Business Media: 329--39.
\href{http://doi.org/10.1007/s10955-012-0596-5}{doi:10.1007/s10955-012-0596-5}.

[19] \hyperdef{}{ref-Domux5fnguezux5fGarcux5faux5f2015}{\label{ref-Domux5fnguezux5fGarcux5faux5f2015}}
inguez-Garcia, Virginia Dom, and Miguel A. Muoz. 2015. ``Ranking Species
in Mutualistic Networks.'' \emph{Sci. Rep.} 5 (February). Nature
Publishing Group: 8182.
\href{http://doi.org/10.1038/srep08182}{doi:10.1038/srep08182}.

[20] \hyperdef{}{ref-Ivanux5f2010}{\label{ref-Ivanux5f2010}}
Ivan, G., and V. Grolmusz. 2010. ``When the Web Meets the Cell: Using
Personalized PageRank for Analyzing Protein Interaction Networks.''
\emph{Bioinformatics} 27 (3). Oxford University Press (OUP): 405--7.
\href{http://doi.org/10.1093/bioinformatics/btq680}{doi:10.1093/bioinformatics/btq680}.

[21] \hyperdef{}{ref-DBLP:confux2fspireux2fJabeurTB12}{\label{ref-DBLP:confux2fspireux2fJabeurTB12}}
Jabeur, Lamjed Ben, Lynda Tamine, and Mohand Boughanem. 2012. ``Active
Microbloggers: Identifying Influencers, Leaders and Discussers in
Microblogging Networks.'' In \emph{String Processing and Information
Retrieval - 19th International Symposium, SPIRE 2012, Cartagena de
Indias, Colombia, October 21-25, 2012. Proceedings}, edited by Liliana
Calder\a'on-Benavides, Cristina N. Gonz\a'alez-Caro, Edgar Ch\a'avez,
and Nivio Ziviani, 7608:111--17. Lecture Notes in Computer Science.
Springer.
\href{http://doi.org/10.1007/978-3-642-34109-0_12}{doi:10.1007/978-3-642-34109-0\_12}.

[22] \hyperdef{}{ref-gleeson2014simple}{\label{ref-gleeson2014simple}}
James P Gleeson, Davide Cellai, Jukka-Pekka Onnela, Mason A Porter, and
Felix Reed-Tsochas. 2014. ``A Simple Generative Model of Collective
Online Behavior.'' \emph{Proceedings of the National Academy of
Sciences} 111 (29). National Acad Sciences: 10411--15.
\href{http://doi.org/10.1073/pnas.1313895111}{doi:10.1073/pnas.1313895111}.

[23] \hyperdef{}{ref-Jiangux5f2008}{\label{ref-Jiangux5f2008}}
Jiang, Bin, Sijian Zhao, and Junjun Yin. 2008. ``Self-Organized Natural
Roads for Predicting Traffic Flow: A Sensitivity Study.'' \emph{J. Stat.
Mech.} 2008 (07). IOP Publishing: P07008.
\href{http://doi.org/10.1088/1742-5468/2008/07/p07008}{doi:10.1088/1742-5468/2008/07/p07008}.

[24] \hyperdef{}{ref-Yushiux5fJingux5f2008}{\label{ref-Yushiux5fJingux5f2008}}
Jing, Yushi, and S. Baluja. 2008. ``VisualRank: Applying PageRank to
Large-Scale Image Search.'' \emph{IEEE Transactions on Pattern Analysis
and Machine Intelligence} 30 (11). Institute of Electrical \&
Electronics Engineers (IEEE): 1877--90.
\href{http://doi.org/10.1109/tpami.2008.121}{doi:10.1109/tpami.2008.121}.

[25] \hyperdef{}{ref-DBLP:journalsux2ftoisux2fHerlockerKTR04}{\label{ref-DBLP:journalsux2ftoisux2fHerlockerKTR04}}
Jonathan L. Herlocker, Joseph A. Konstan, Loren G. Terveen, and John
Riedl. 2004. ``Evaluating Collaborative Filtering Recommender Systems.''
\emph{ACM Trans. Inf. Syst.} 22 (1): 5--53.
\href{http://doi.org/10.1145/963770.963772}{doi:10.1145/963770.963772}.

[26] \hyperdef{}{ref-DBLP:journalsux2ftwebux2fLeskovecAH07}{\label{ref-DBLP:journalsux2ftwebux2fLeskovecAH07}}
Jure Leskovec, Lada A. Adamic, and Bernardo A. Huberman. 2007. ``The
Dynamics of Viral Marketing.'' \emph{TWEB} 1 (1).
\href{http://doi.org/10.1145/1232722.1232727}{doi:10.1145/1232722.1232727}.

[27] \hyperdef{}{ref-LeskovecBK09}{\label{ref-LeskovecBK09}}
Jure Leskovec, Lars Backstrom, and Jon M. Kleinberg. 2009.
``Meme-Tracking and the Dynamics of the News Cycle.'' In
\emph{Proceedings of the 15th ACM SIGKDD International Conference on
Knowledge Discovery and Data Mining, Paris, France, June 28 - July 1,
2009}, edited by John F. Elder IV, Françoise Fogelman-Souli\a'e, Peter
A. Flach, and Mohammed Javeed Zaki, 497--506. ACM.
\href{http://doi.org/10.1145/1557019.1557077}{doi:10.1145/1557019.1557077}.

[28] \hyperdef{}{ref-Kempeux5f2003}{\label{ref-Kempeux5f2003}}
Kempe, David, Jon Kleinberg, and \a'Eva Tardos. 2003. ``Maximizing the
Spread of Influence Through a Social Network.'' In \emph{Proceedings of
the Ninth ACM SIGKDD International Conference on Knowledge Discovery and
Data Mining - KDD 03}. Association for Computing Machinery (ACM).
\href{http://doi.org/10.1145/956750.956769}{doi:10.1145/956750.956769}.

[29] \hyperdef{}{ref-DBLP:confux2fsocialcomux2fKhrabrovC10}{\label{ref-DBLP:confux2fsocialcomux2fKhrabrovC10}}
Khrabrov, Alexy, and George Cybenko. 2010. ``Discovering Influence in
Communication Networks Using Dynamic Graph Analysis.'' In
\emph{Proceedings of the 2010 IEEE Second International Conference on
Social Computing, SocialCom / IEEE International Conference on Privacy,
Security, Risk and Trust, PASSAT 2010, Minneapolis, Minnesota, USA,
August 20-22, 2010}, edited by Ahmed K. Elmagarmid and Divyakant
Agrawal, 288--94. IEEE Computer Society.
\href{http://doi.org/10.1109/SocialCom.2010.48}{doi:10.1109/SocialCom.2010.48}.

[30] \hyperdef{}{ref-DBLP:journalsux2fcorrux2fKhushnoodSL16}{\label{ref-DBLP:journalsux2fcorrux2fKhushnoodSL16}}
Khushnood, Abbas, Mingsheng Shang, and Xin Luo. 2016. ``Discovering
Items with Potential Popularity on Social Media.'' \emph{CoRR}
abs/1604.01131. \url{http://arxiv.org/abs/1604.01131}.

[31] \hyperdef{}{ref-Lambiotteux5f2005}{\label{ref-Lambiotteux5f2005}}
Lambiotte, R., and M. Ausloos. 2005. ``Uncovering Collective Listening
Habits and Music Genres in Bipartite Networks.'' \emph{Physical Review
E} 72 (6). American Physical Society (APS).
\href{http://doi.org/10.1103/physreve.72.066107}{doi:10.1103/physreve.72.066107}.

[32] \hyperdef{}{ref-Lappasux5f2010}{\label{ref-Lappasux5f2010}}
Lappas, Theodoros, Evimaria Terzi, Dimitrios Gunopulos, and Heikki
Mannila. 2010. ``Finding Effectors in Social Networks.'' In
\emph{Proceedings of the 16th ACM SIGKDD International Conference on
Knowledge Discovery and Data Mining - KDD 10}. Association for Computing
Machinery (ACM).
\href{http://doi.org/10.1145/1835804.1835937}{doi:10.1145/1835804.1835937}.

[33] \hyperdef{}{ref-DBLP:confux2fasunamux2fLiCCJ13}{\label{ref-DBLP:confux2fasunamux2fLiCCJ13}}
Li, Xiang, Shaoyin Cheng, Wenlong Chen, and Fan Jiang. 2013. ``Novel
User Influence Measurement Based on User Interaction in Microblog.'' In
\emph{Advances in Social Networks Analysis and Mining 2013, ASONAM '13,
Niagara, oN, Canada - August 25 - 29, 2013}, edited by Jon G. Rokne and
Christos Faloutsos, 615--19. ACM.
\href{http://doi.org/10.1145/2492517.2492635}{doi:10.1145/2492517.2492635}.

[34] \hyperdef{}{ref-greycite40361}{\label{ref-greycite40361}}
louridas. 2016. ``Louridas/PageRank.'' \emph{GitHub}.
\url{https://github.com/louridas/PageRank}.
\url{https://github.com/louridas/PageRank}.

[35] \hyperdef{}{ref-Maux5f2008}{\label{ref-Maux5f2008}}
Ma, Nan, Jiancheng Guan, and Yi Zhao. 2008. ``Bringing PageRank to the
Citation Analysis.'' \emph{Information Processing \& Management} 44 (2).
Elsevier BV: 800--810.
\href{http://doi.org/10.1016/j.ipm.2007.06.006}{doi:10.1016/j.ipm.2007.06.006}.

[36] \hyperdef{}{ref-DBLP:confux2fsofsemux2fMajerS12}{\label{ref-DBLP:confux2fsofsemux2fMajerS12}}
Majer, Tomas, and Marian Simko. 2012. ``Leveraging Microblogs for
Resource Ranking.'' In \emph{SOFSEM 2012: Theory and Practice of
Computer Science - 38th Conference on Current Trends in Theory and
Practice of Computer Science, Špindleruv Mlýn, Czech Republic, January
21-27, 2012. Proceedings}, edited by M\a'aria Bielikov\a'a, Gerhard
Friedrich, Georg Gottlob, Stefan Katzenbeisser, and György Tur\a'an,
7147:518--29. Lecture Notes in Computer Science. Springer.
\href{http://doi.org/10.1007/978-3-642-27660-6_42}{doi:10.1007/978-3-642-27660-6\_42}.

[37] \hyperdef{}{ref-Medoux5f2011}{\label{ref-Medoux5f2011}}
Medo, Mat\a'uš, Giulio Cimini, and Stanislao Gualdi. 2011. ``Temporal
Effects in the Growth of Networks.'' \emph{Phys. Rev. Lett.} 107 (23).
American Physical Society (APS).
\href{http://doi.org/10.1103/physrevlett.107.238701}{doi:10.1103/physrevlett.107.238701}.

[38] \hyperdef{}{ref-Moroneux5f2015}{\label{ref-Moroneux5f2015}}
Morone, Flaviano, and Hern\a'an A. Makse. 2015. ``Influence Maximization
in Complex Networks Through Optimal Percolation.'' \emph{Nature} 524
(7563). Nature Publishing Group: 65--68.
\href{http://doi.org/10.1038/nature14604}{doi:10.1038/nature14604}.

[39] \hyperdef{}{ref-Pastorux5fSatorrasux5f2002}{\label{ref-Pastorux5fSatorrasux5f2002}}
Pastor-Satorras, Romualdo, and Alessandro Vespignani. 2002.
``Immunization of Complex Networks.'' \emph{Physical Review E} 65 (3).
American Physical Society (APS).
\href{http://doi.org/10.1103/physreve.65.036104}{doi:10.1103/physreve.65.036104}.

[40] \hyperdef{}{ref-Piccardiux5f2008}{\label{ref-Piccardiux5f2008}}
Piccardi, Carlo, and Renato Casagrandi. 2008. ``Inefficient Epidemic
Spreading in Scale-Free Networks.'' \emph{Physical Review E} 77 (2).
American Physical Society (APS).
\href{http://doi.org/10.1103/physreve.77.026113}{doi:10.1103/physreve.77.026113}.

[41] \hyperdef{}{ref-DBLP:journalsux2fjoiux2fParoloPGHKF15}{\label{ref-DBLP:journalsux2fjoiux2fParoloPGHKF15}}
Pietro Della Briotta Parolo, Raj Kumar Pan, Rumi Ghosh, Bernardo A.
Huberman, Kimmo Kaski, and Santo Fortunato. 2015. ``Attention Decay in
Science.'' \emph{J. Informetrics} 9 (4): 734--45.
\href{http://doi.org/10.1016/j.joi.2015.07.006}{doi:10.1016/j.joi.2015.07.006}.

[42] \hyperdef{}{ref-Richardsonux5f2002}{\label{ref-Richardsonux5f2002}}
Richardson, Matthew, and Pedro Domingos. 2002. ``Mining
Knowledge-Sharing Sites for Viral Marketing.'' In \emph{Proceedings of
the Eighth ACM SIGKDD International Conference on Knowledge Discovery
and Data Mining - KDD 02}. Association for Computing Machinery (ACM).
\href{http://doi.org/10.1145/775047.775057}{doi:10.1145/775047.775057}.

[43] \hyperdef{}{ref-Sayyadiux5f2009}{\label{ref-Sayyadiux5f2009}}
Sayyadi, Hassan, and Lise Getoor. 2009. ``FutureRank: Ranking Scientific
Articles by Predicting Their Future PageRank.'' In \emph{Proceedings of
the 2009 SIAM International Conference on Data Mining}, 533--44. Society
for Industrial \& Applied Mathematics (SIAM).
\href{http://doi.org/10.1137/1.9781611972795.46}{doi:10.1137/1.9781611972795.46}.

[44] \hyperdef{}{ref-Shangux5f2010}{\label{ref-Shangux5f2010}}
Shang, Ming-Sheng, Linyuan Lü, Yi-Cheng Zhang, and Tao Zhou. 2010.
``Empirical Analysis of Web-Based User-Object Bipartite Networks.''
\emph{EPL} 90 (4). IOP Publishing: 48006.
\href{http://doi.org/10.1209/0295-5075/90/48006}{doi:10.1209/0295-5075/90/48006}.

[45] \hyperdef{}{ref-Subbianux5f2016}{\label{ref-Subbianux5f2016}}
Subbian, Karthik, Charu Aggarwal, and Jaideep Srivastava. 2016. ``Mining
Influencers Using Information Flows in Social Streams.'' \emph{ACM
Trans. Knowl. Discov. Data} 10 (3). Association for Computing Machinery
(ACM): 1--28.
\href{http://doi.org/10.1145/2815625}{doi:10.1145/2815625}.

[46] \hyperdef{}{ref-Subbianux5f2011}{\label{ref-Subbianux5f2011}}
Subbian, Karthik, and Prem Melville. 2011. ``Supervised Rank Aggregation
for Predicting Influencers in Twitter.'' In \emph{2011 IEEE Third Intl
Conference on Privacy, Security, Risk and Trust and 2011 IEEE Third Intl
Conference on Social Computing}. Institute of Electrical \& Electronics
Engineers (IEEE).
\href{http://doi.org/10.1109/passat/socialcom.2011.167}{doi:10.1109/passat/socialcom.2011.167}.

[47] \hyperdef{}{ref-Wangux5f2016}{\label{ref-Wangux5f2016}}
Wang, Zhixiao, Ya Zhao, Jingke Xi, and Changjiang Du. 2016. ``Fast
Ranking Influential Nodes in Complex Networks Using a K-Shell Iteration
Factor.'' \emph{Physica A: Statistical Mechanics and Its Applications}
461 (November). Elsevier BV: 171--81.
\href{http://doi.org/10.1016/j.physa.2016.05.048}{doi:10.1016/j.physa.2016.05.048}.

[48] \hyperdef{}{ref-DBLP:confux2fwsdmux2fWengLJH10}{\label{ref-DBLP:confux2fwsdmux2fWengLJH10}}
Weng, Jianshu, Ee-Peng Lim, Jing Jiang, and Qi He. 2010. ``TwitterRank:
Finding Topic-Sensitive Influential Twitterers.'' In \emph{Proceedings
of the Third International Conference on Web Search and Web Data Mining,
WSDM 2010, New York, NY, USA, February 4-6, 2010}, edited by Brian D.
Davison, Torsten Suel, Nick Craswell, and Bing Liu, 261--70. ACM.
\href{http://doi.org/10.1145/1718487.1718520}{doi:10.1145/1718487.1718520}.

[49] \hyperdef{}{ref-Wuux5f2010}{\label{ref-Wuux5f2010}}
Wu, Gang, and Yimin Wei. 2010. ``Arnoldi Versus GMRES for Computing
PageRank.'' \emph{ACM Transactions on Information Systems} 28 (3).
Association for Computing Machinery (ACM): 1--28.
\href{http://doi.org/10.1145/1777432.1777434}{doi:10.1145/1777432.1777434}.

[50] \hyperdef{}{ref-Yanux5f2006}{\label{ref-Yanux5f2006}}
Yan, Gang, Tao Zhou, Bo Hu, Zhong-Qian Fu, and Bing-Hong Wang. 2006.
``Efficient Routing on Complex Networks.'' \emph{Physical Review E} 73
(4). American Physical Society (APS).
\href{http://doi.org/10.1103/physreve.73.046108}{doi:10.1103/physreve.73.046108}.

[51] \hyperdef{}{ref-Zhouux5f2015}{\label{ref-Zhouux5f2015}}
Yanbo Zhou, An Zeng, and Wei-Hong Wang. 2015. ``Temporal Effects in
Trend Prediction: Identifying the Most Popular Nodes in the Future.''
Edited by Xia Li. \emph{PLOS ONE} 10 (3). Public Library of Science
(PLoS): e0120735.
\href{http://doi.org/10.1371/journal.pone.0120735}{doi:10.1371/journal.pone.0120735}.

[52] \hyperdef{}{ref-DBLP:confux2fwsdmux2fYangL11}{\label{ref-DBLP:confux2fwsdmux2fYangL11}}
Yang, Jaewon, and Jure Leskovec. 2011. ``Patterns of Temporal Variation
in Online Media.'' In \emph{Proceedings of the Forth International
Conference on Web Search and Web Data Mining, WSDM 2011, Hong Kong,
China, February 9-12, 2011}, edited by Irwin King, Wolfgang Nejdl, and
Hang Li, 177--86. ACM.
\href{http://doi.org/10.1145/1935826.1935863}{doi:10.1145/1935826.1935863}.

\end{document}